\documentclass[amssymb,amsmath,preprint]{revtex4-1}

\usepackage{graphicx}
\usepackage{bm}
\usepackage{epsfig}

\newcommand{\nsub}[1]{_{\rm{#1}}}
\newcommand{\nsup}[1]{^{\rm{#1}}}
\newcommand{\grad}{\ensuremath{^\circ}}

\begin{document}

\title{The new Cold Neutron Chopper Spectrometer at the Spallation Neutron Source -- Design and Performance}

\author{G. Ehlers}
\affiliation{Neutron Scattering Science Division, Oak Ridge National
Laboratory, Oak Ridge, TN 37831, USA}
\author{A. A. Podlesnyak}
\affiliation{Neutron Scattering Science Division, Oak Ridge National
Laboratory, Oak Ridge, TN 37831, USA}
\author{J. L. Niedziela}
\affiliation{Neutron Scattering Science Division, Oak Ridge National
Laboratory, Oak Ridge, TN 37831, USA}
\author{E. B. Iverson}
\affiliation{Neutron Scattering Science Division, Oak Ridge National
Laboratory, Oak Ridge, TN 37831, USA}
\author{P. E. Sokol}
\affiliation{Department of Physics, Indiana University, Bloomington, IN 47405, USA}

\begin{abstract}
The design and performance of the new Cold Neutron Chopper Spectrometer (CNCS) at the Spallation Neutron Source in Oak Ridge are described.
CNCS is a direct geometry inelastic time-of-flight spectrometer, designed to cover essentially the same energy and momentum transfer ranges as IN5 at ILL, LET at ISIS, DCS at NIST, TOFTOF at FRM2, AMATERAS at J-PARC, PHAROS at LANSCE and NEAT at HZB, at similar energy resolution.
Measured values of key figures such as neutron flux at sample position and energy resolution are compared between measurements and ray tracing Monte Carlo simulations, and good agreement (better than 20\% of absolute numbers) has been achieved.
The instrument performs very well in the cold and thermal neutron energy ranges, and promises to become a workhorse for the neutron scattering community for quasielastic and inelastic scattering experiments.
\end{abstract}

\preprint{Journal-ref: \textit{Review of Scientific Instruments} \textbf{82}, 085108 (2011)}
\maketitle

\section{Introduction}
\label{Intro}

The Cold Neutron Chopper Spectrometer (CNCS) at the Spallation Neutron Source (SNS) in Oak Ridge is a very flexible and versatile direct-geometry multi-chopper inelastic time-of-flight (TOF) spectrometer that provides both good energy and momentum transfer resolution at low incident neutron energies (1-50 meV).
This type of instrument is very popular at pulsed as well as continuous sources for the wide range of scientific applications
offered~\cite{Copley,in5,isislet,toftof,amateras,pharos,neat}.
Therefore, most of the major neutrons sources around the world host at least one instrument that is closely comparable to CNCS, such as IN5 at ILL, LET at ISIS, DCS at NIST, TOFTOF at FRM2, AMATERAS at J-PARC, PHAROS at LANSCE and NEAT at HZB (formerly HMI Berlin).
The CNCS is located at SNS beam line 5 with a cold coupled moderator, and features a source to sample distance of 36.25~m~\cite{Mason06}.
Two high-speed choppers, one to shape the neutron pulse from the moderator and a second to cut down the pulse length at the sample position, provide an adjustable energy resolution, ranging from $\sim{1.2}$\% to $\sim{10}$\% of the incident energy.
The secondary flight path, with a length of 3.5 m, has a highly pixilated detector covering scattering angles between $-50$\grad\ and $+135$\grad\ in the scattering plane and $\pm{16}$\grad\ perpendicular to the scattering plane.
The detector array with a total solid angle of 1.7 sr consists of 400 two meter long tubes filled with $\nsup{3}$He gas at 6 atm.
CNCS has been operating since May 2009 (more than sixty peer-reviewed user experiments to date) with external facility users.

\section{Basic Design Choices}
\label{Design}

\begin{figure}[t]
\includegraphics[width=6.0in]{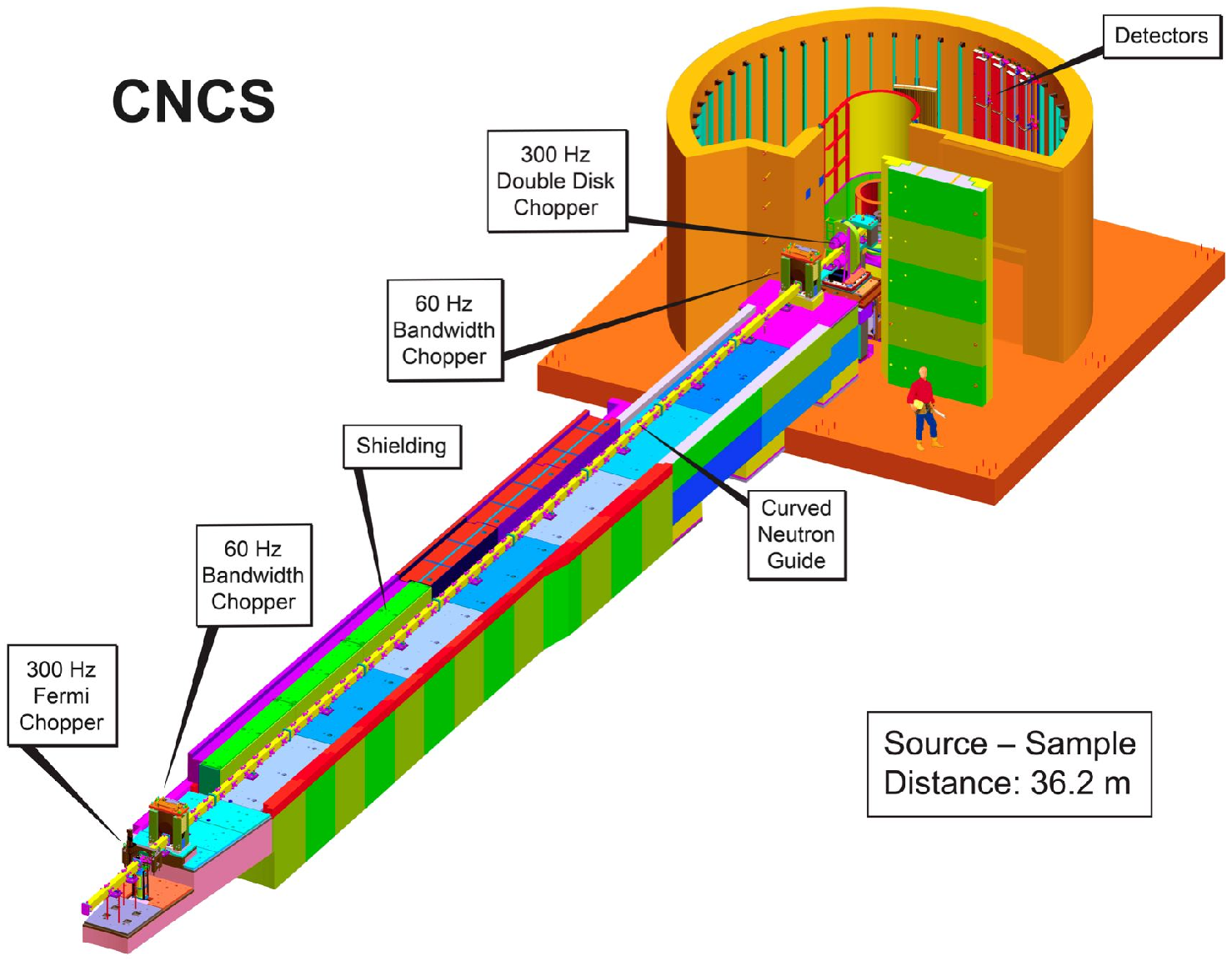}
\caption{\label{illu} General layout of CNCS. Instrument shielding is shown in part, to reveal chopper and guide locations. Likewise, the roof of the detector enclosure has been removed to show the detectors.}
\end{figure}

The general layout of the instrument is shown in Fig.~\ref{illu}.
The chopper system consists of four choppers~\cite{Sokol}.
Going downstream, the first chopper is a pulse-shaping Fermi chopper capable of spinning at up to 300 Hz.
Two slit packages are mounted on the same axis of rotation and can be changed with a vertical translation mechanism at full rotational speed.
The slit packages are relatively short (17~mm) with a shallow curvature of the slits, the overall neutron transmission of the device is $\sim{85}$\% when facing the beam.
The second and third choppers are 60~Hz disk choppers designed to remove frame overlap from the neutron beam.
These choppers have one opening of 14.3\grad\ cut in the disks.
The final chopper, at a distance of 1.5~m from the sample, is a high-speed (300 Hz) chopper with two counter-rotating disks.
Each disk has three slits with widths of 2\grad, 4.4\grad\ and 9\grad, respectively.
These openings can be paired in various ways to provide flexibility for the energy resolution setting.
All choppers can be moved out of the beam or stopped `open' to allow for `white beam' operation.
The two high speed choppers determine the neutron energy at which the instrument operates via the relative phase between them, and the energy resolution through the chopper opening times.
Even though the moderator pulse is relatively short (generally $\lesssim{300}{\,}\mu\rm{s}$, depending on neutron energy), it is generally too long for most applications. The pulse-shaping Fermi chopper offers flexibility in varying the effectively used moderator pulse length.
The chopper vacuum is separated from the guide vacuum for each individual chopper.

The overall design of the neutron guide was optimized for neutron flux at the sample position, at 25~cm behind the guide exit, for a sample shaped like the most commonly used cylindrical containers.
The design optimization was achieved in an iterative process, using the McStas
package~\cite{Lefmann,McStas}, while maintaining a few boundary conditions:
\begin{itemize}
\item a curvature is needed to reduce high-energy and thermal neutron background
\item the overall length impacts the resulting energy resolution of the instrument and must match other contributions to the resolution
\item a structural column of the target building had to be avoided
\item it was desirable to maintain a `simple' design to keep the manufacturing and installation cost within reasonable limits
\item it was highly desirable to have a narrow beam at the location of the last chopper to help achieve the shortest pulse possible at the sample.
\end{itemize}
The neutron guide entrance is located at a distance of 1~m from the moderator surface and has an initial cross section of 100~mm $\times$ 50~mm (height $\times$ width).
The overall length of the guide is 34.95~m.
The first section of the guide is straight, over a length of 6.57~m.  This is followed by a curved section (radius of curvature 2000~m) of 15~m length.
The remainder of the guide is straight again.
The curvature ensures that the guide exit is out of direct line-of-sight from the guide entrance, effectively introducing a high-energy cut-off at
around $\sim{100}$~meV in the neutron transmission of the guide.
At the end, over a length of 5.19~m, the guide is tapered for a cross section at the exit of 50~mm $\times$ 15~mm (height $\times$ width).
The guide is coated with supermirrors, mostly $m=2.5$, but increasing to $m=4$ towards the end.

Two low-efficiency beam monitors are placed in the incident beam line.
The first monitor is situated downstream of the first frame overlap chopper.
The second of the monitors is placed directly behind the high speed sample chopper and is calibrated for efficiency, allowing approximate estimates of the neutron flux at this position in the beam.

The sample area is designed to allow for bulky equipment, such as cryogenic magnets with a diameter of up to 1~m, to be placed on the sample table.
The last (relatively short) section of the neutron guide can be removed, putting the guide exit at 55~cm from the sample position.
In standard configuration, a 2\grad\ radial collimator is mounted to the table to suppress scattering from the sample environment.

Detectors are situated at 3.5~m distance from the sample inside the detector enclosure.
During normal operation the enclosure is filled with argon gas, with the addition of 2\% CO$_2$ to increase the breakdown voltage of the argon.
Detectors are conventional $\nsup{3}$He tubes which are position sensitive along an active length of $\sim{1.95}$~m, operating at 1650~V with
6~atm $\nsup{3}$He gas pressure.
A total of 400 tubes cover scattering angles between $-50$\grad\ and $+135$\grad\ in the scattering plane and $\pm{16}$\grad\ perpendicular to the scattering plane, resulting in a total solid angle of 1.7~sr.
During neutron beam production the overall detector background (electronic noise and background neutrons) amounts to less than 0.5 counts per minute per meter of tube.

\section{Instrument Performance}
\label{Performance}

\begin{figure}[t]
\includegraphics[width=5.0in]{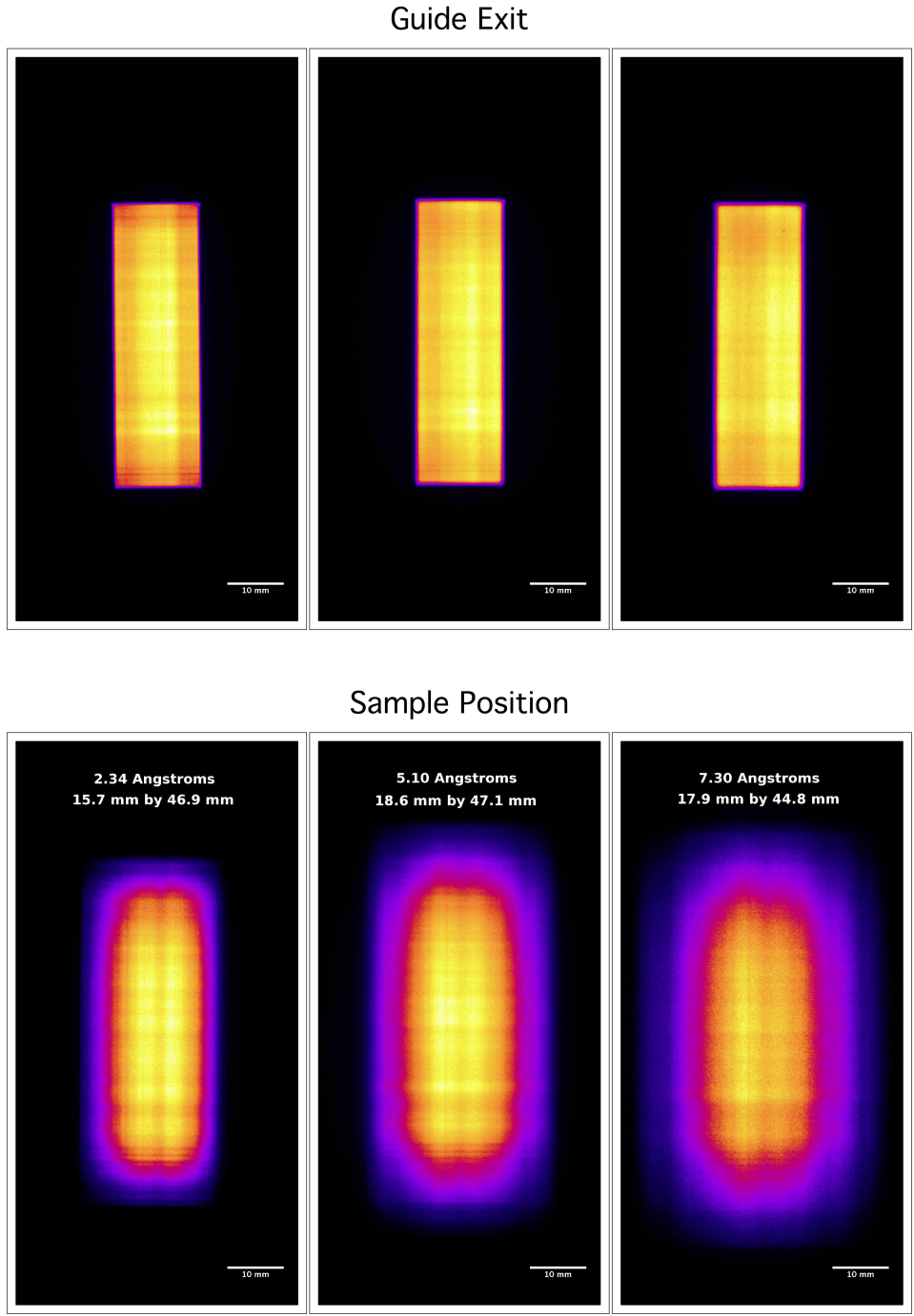}
\caption{\label{hirespic} High resolution images of the CNCS beam at the guide exit (top) and sample position (bottom). In the lower part of the figure, the beam edges are taken at 50\% intensity maximum.}
\end{figure}

High resolution images of the beam were taken at the guide exit and at the sample position.
These images are shown in Fig.~\ref{hirespic}.
Due to the wavelength dependent beam divergence, the beam at the sample position is generally somewhat larger than at the guide exit.

The instrument performance in terms of flux and resolution was characterized with various measurements.
These measurements are compared with the results of ray tracing Monte Carlo simulations.
The simulations were performed with the McStas
package~\cite{Lefmann,McStas} and are based on a source term that was developed for the cold coupled SNS moderator~\cite{Eriks}.
The goal was to assess whether all CNCS components work in optimal regime with no substantial loss of instrumental resolution/intensity.
The model was made up of the following components: the SNS neutron source that produces a time and energy distribution from the SNS top-downstream cold coupled moderator; 11 individual neutron guide sections (component ``Guide'') including 15 meters of a curved section (component ``Bender''),
Fermi chopper (component ``Vitess{\_}ChopperFermi''); two band-width disk choppers and one double disk chopper (component ``DiskChopper''); radial collimator; detectors (component ``Monitor{\_}nD''); Al windows; beam monitors.
The customized radial collimator module was written using the ``Exact{\_}radial{\_}coll'' component~\cite{McStas} with rectangular openings, specified length and blade thickness.
The number of simulated neutrons was $1\times{10}^9$ for each wavelength, which took about 30 minutes using parallel computing (MPI).
The simulations are showing good agreement between the calculated and measured data in a wide range of incident energies.
The absolute intensities as well as TOF peak-width at both the calibrated downstream monitor and the sample position were reproduced with reasonable accuracy, typically better than 20\%.

\begin{figure}[t]
\includegraphics[width=6.0in]{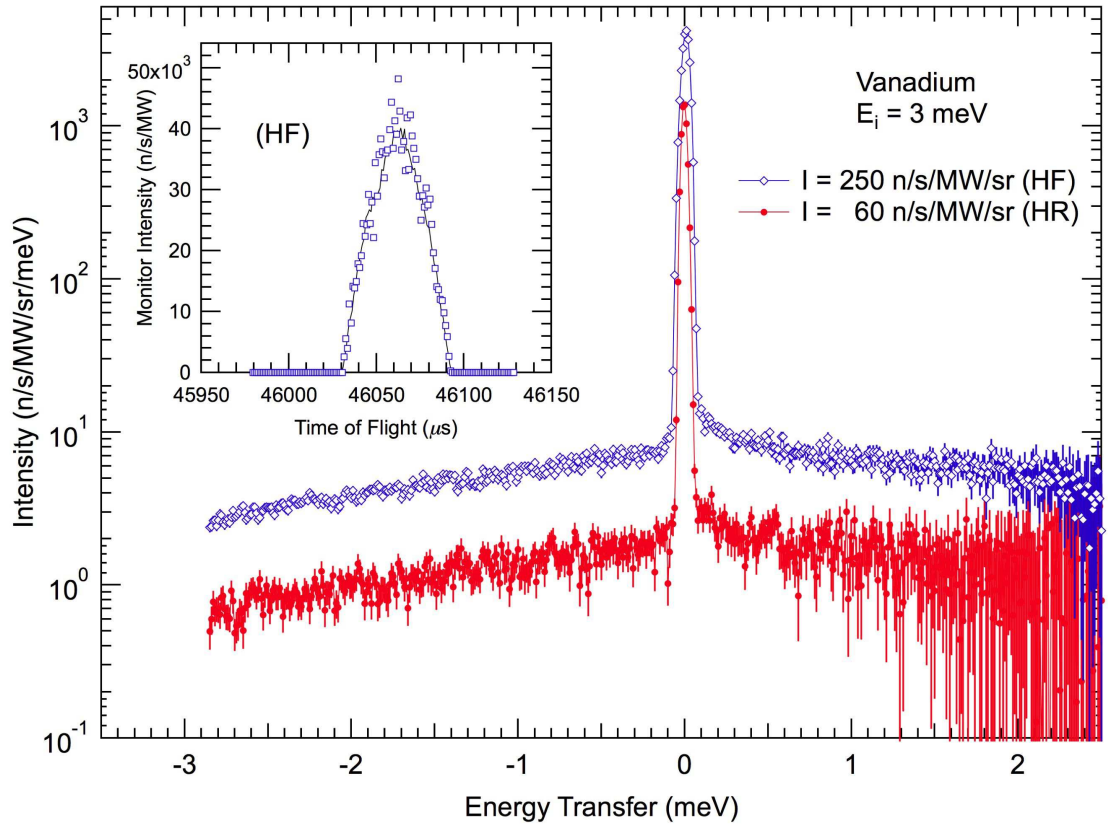}
\caption{\label{vana} Main figure: detector intensity on a standard vanadium sample normalized to time, source power and detector area (raw data converted to $S(\omega)$). Inset: Intensity in the calibrated monitor (HF setting) at the same energy, data and McStas simulation (solid line). Intensities are in absolute units. The peak to background ratio of $10^3$ is routinely achieved.}
\end{figure}

\begin{figure}[t]
\includegraphics[width=6.0in]{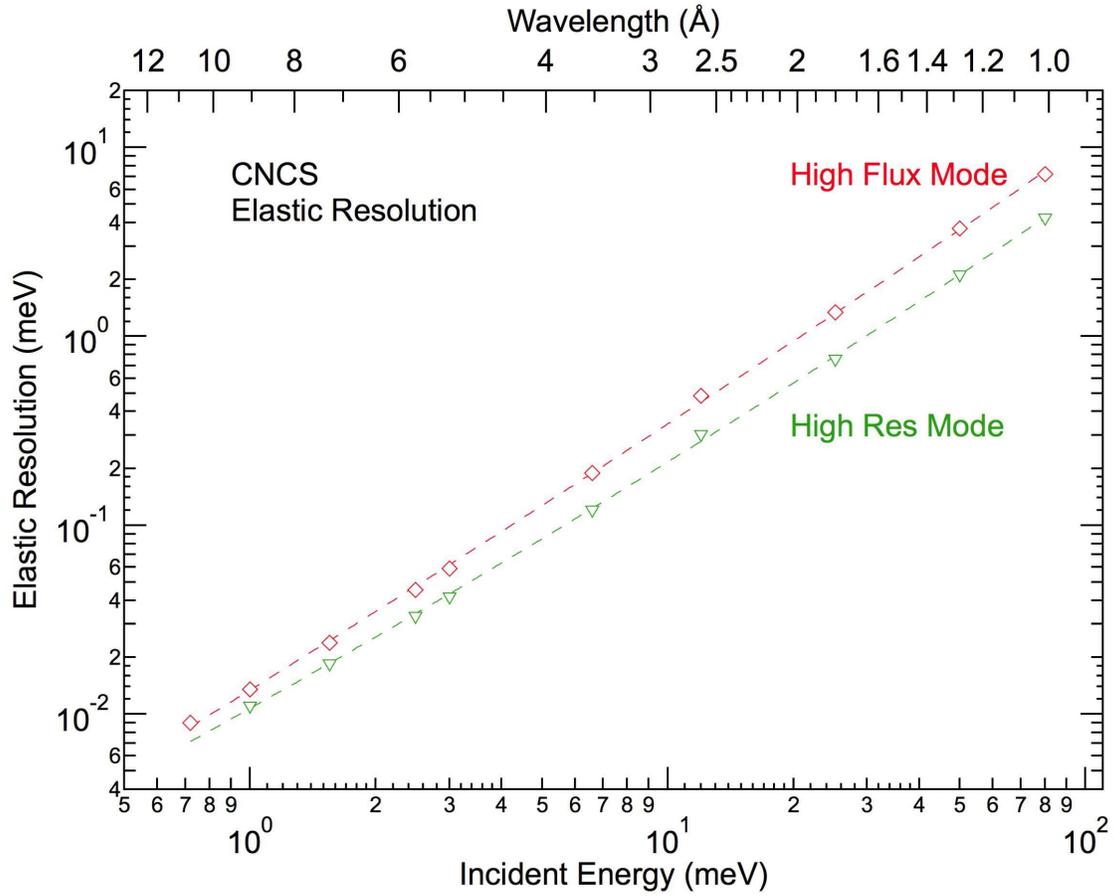}
\caption{\label{elres} Elastic energy resolution measured for two standard settings ($\diamond$ and $\triangledown$) with a standard scatterer (vanadium cylinder). Lines are analytically calculated as described in the text.}
\end{figure}

Measurements of the overall intensity and elastic instrument resolution were made with a standard vanadium sample, a solid cylinder with a diameter of 6.35~mm and a length of 5~cm.
This is shown in Figs.~\ref{vana} and \ref{elres}.
These data were complemented by measurements of the neutron flux at the sample position using two calibrated monitors.
While the instrument allows for many different settings for the resolution at a given incident energy, by changing the speed of the Fermi chopper and the sample chopper, and by pairing different slit combinations of the sample chopper disks, in standard operation most often one of two modes is chosen.
For convenience, these modes are termed `high resolution' (HR) and `high flux' (HF).
The `high resolution' setting gives nearly the best achievable resolution.
The `high flux' setting results in a neutron flux at the sample position that is generally about 4 times higher, at a somewhat relaxed resolution
(see Fig.~\ref{elres}).
The energy resolution can be further relaxed for more intensity.
In both standard modes the Fermi chopper runs at 180 Hz and the sample chopper runs at 300 Hz, and the two modes differ in the choice of the sample chopper slits that are paired.
At $E\nsub{i}=3$~meV ($\lambda\nsub{i}=5.2$~{\AA}), the elastic resolution for the HF and HR settings is 59~$\mu$eV and 42~$\mu$eV, respectively, and the measured flux at the sample (full beam cross section, flux normalized to a source power of 1~MW) is $7.6\times{10}^5$~n/s/MW and
$2.6\times{10}^5$~n/s/MW, respectively.
The best resolution comes at a price in terms of intensity, because a gain in resolution by a factor of $59/42\sim{1.4}$ is in general not quite worth an intensity loss by a factor of $7.6/2.6\sim{3}$.
This observation is also made at other operating energies.

\begin{figure}[t]
\includegraphics[width=6.0in]{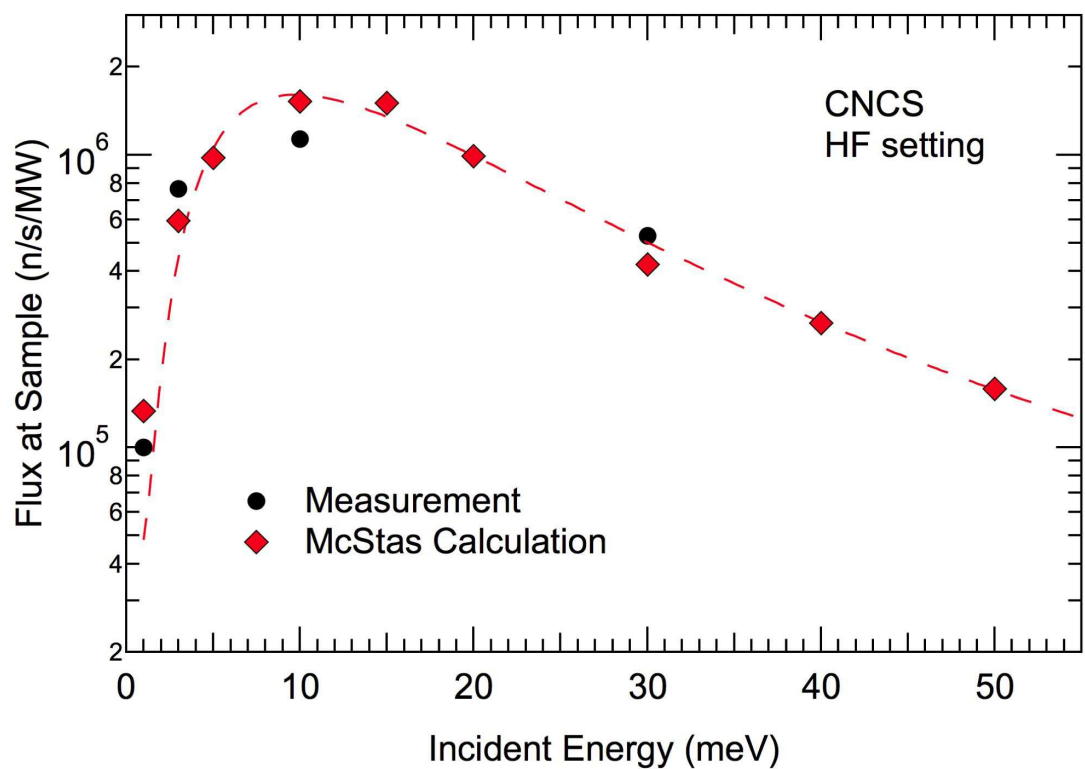}
\caption{\label{flux} Sample flux (full beam) in the HF setting at different incident energies. The peak flux is obtained around 10 meV. The line is a guide to the eye.}
\end{figure}

The dependence of the sample flux on the incident energy is shown in
Fig.~\ref{flux}.
The measurements show that the peak flux is obtained around
$E\nsub{i}=10$~meV ($\lambda=2.9$~{\AA}).
Thus not only cold but also thermal neutron energies are readily available at CNCS.

\begin{figure}[t]
\includegraphics[width=6.0in]{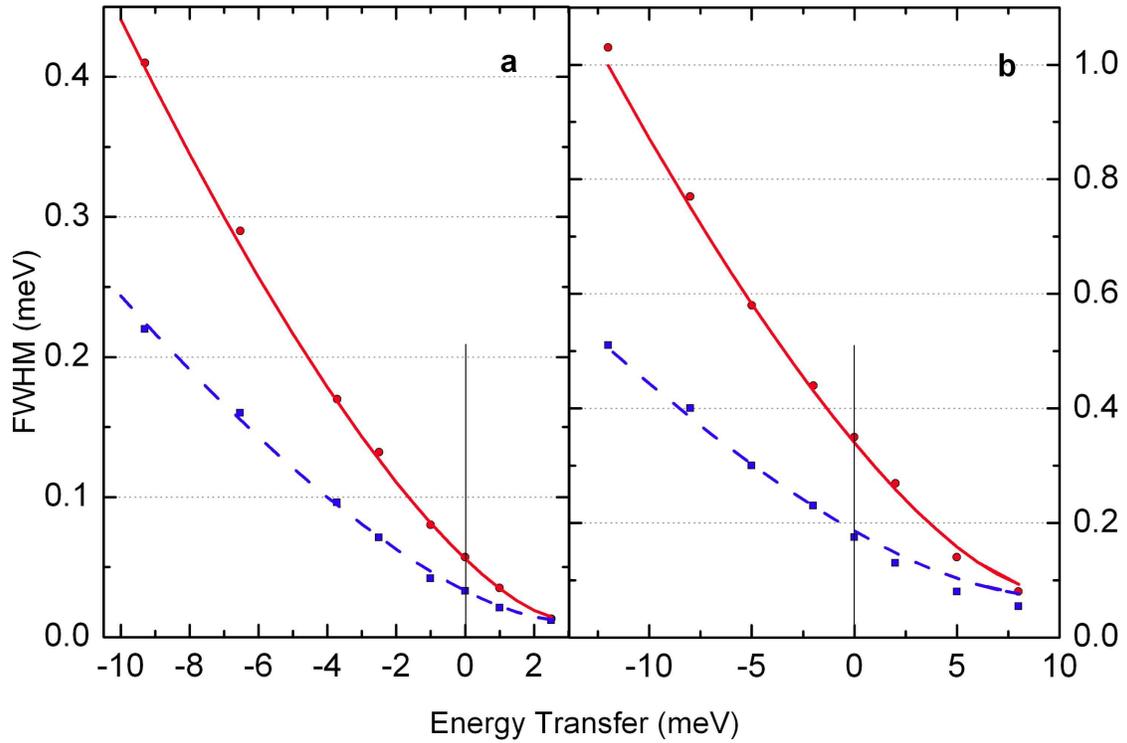}
\caption{\label{inelreso} McStas calculation of the inelastic energy resolution for two popular instrument settings with (a) $E\nsub{i}=3$~meV
and (b) $E\nsub{i}=10$~meV in both HF and HR mode.
These calculations were done for a typical cylindrical sample (40 mm height, 12 mm diameter). Lines are analytically calculated as described in the text. }
\end{figure}

\begin{figure}[t]
\includegraphics[width=6.0in]{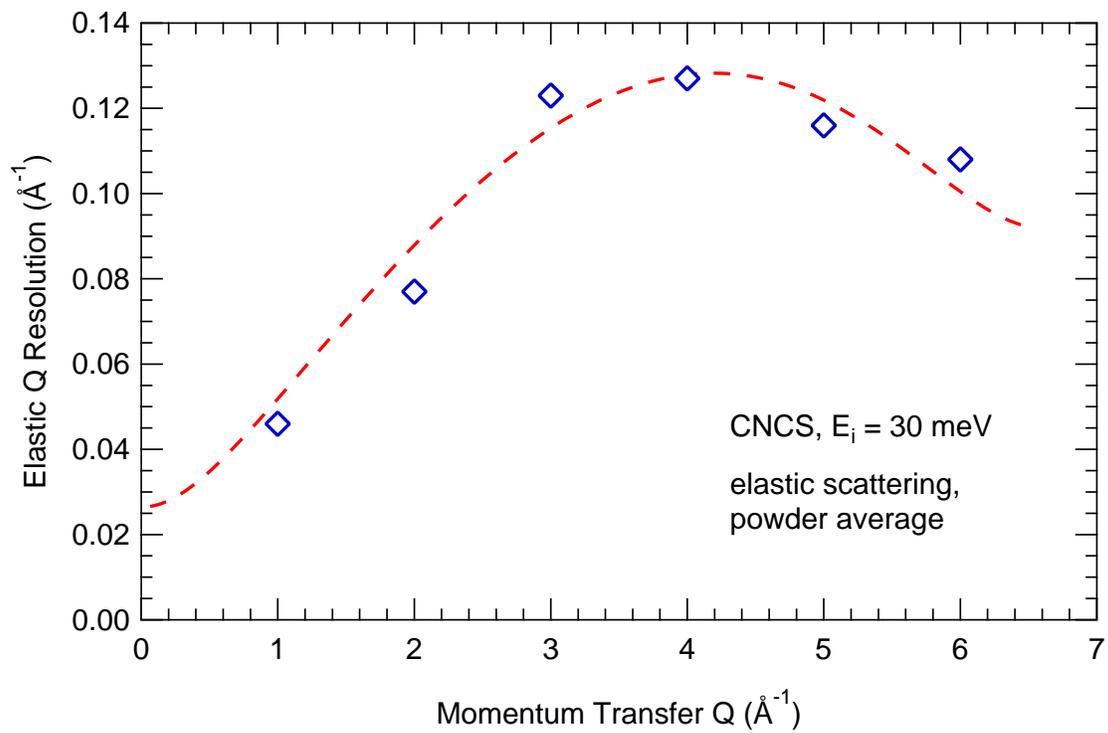}
\caption{\label{momreso} McStas calculation of the elastic momentum transfer resolution (powder average) for $E\nsub{i}=30$~meV in HF mode.
These calculations were done for a typical cylindrical sample (40 mm height, 12 mm diameter). Lines are analytically calculated as described in the text. }
\end{figure}

The performance evaluation presented here concludes with a discussion of the energy and $Q$ resolution as computed with the McStas model,
see Figs.~\ref{inelreso} and \ref{momreso}.
The energy resolution of a direct geometry chopper instrument is given by
\begin{displaymath}
\delta{E} = m_n\cdot\left\{\left(\frac{v_{i}^{3}}{L_{1}}+
\frac{v_f^{3}L_{2}}{L_{1}L_{3}}\right)^{{\!}2}{\!}\delta{t}_{p}^{2}+
\left(\frac{v_{i}^{3}}{L_{1}}+\frac{v_f^{3}(L_{1}+L_{2})}{L_{1}L_{3}}
\right)^{{\!}2}{\!}\delta{t}_{c}^{2}+
\left(\frac{v_{f}^{3}}{L_{3}}\right)^{{\!}2}{\!}\delta{t}_{d}^{2}
\right\}^{1/2}
\end{displaymath}
where $E=m_n{/2}\cdot\left({v_{i}^{2}-v_{f}^{2}}\right)$ is the neutron energy transfer, $m_n$ is the neutron mass, $v_i$ and $v_f$ are the initial and final neutron velocities, respectively, $L_1$ is the distance from the pulse shaping chopper to the sample chopper, $L_2$ is the distance from the sample chopper to the sample, $L_3$ is the distance from the sample to the detector,
$\delta{t}_{p}$ is the pulse width at the pulse shaping chopper,
$\delta{t}_{c}$ is the pulse width at the sample chopper, and
$\delta{t}_{d}$ is the uncertainty of the time-of-flight in the secondary spectrometer due to the physical extension of the sample and the neutron detection volume~\cite{ANLreport}.
The $Q$ resolution can be expressed likewise, and since most of the terms are angle dependent, it is convenient to split the vector $Q$ in two components, which are parallel ($Q_x$) and perpendicular ($Q_y$) to the incident beam direction, respectively. These components read
\begin{eqnarray*}
Q_{x} & = & \frac{m_{n}}{\hbar}\cdot\left[v_{i}-v_{f}\cdot\cos{2}\theta\right] \\[1em]
Q_{y} & = & \frac{m_{n}}{\hbar}\cdot\left[-v_{f}\cdot\sin{2}\theta\right] {\;}.
\end{eqnarray*}
The $Q$ resolution is then
\begin{displaymath}
\delta{Q}=\frac{1}{Q}\cdot\left[Q_{x}^{2}\left(\delta{Q}_{x}\right)^{2}+
Q_{y}^{2}\left(\delta{Q}_{y}\right)^{2}\right]^{1/2} {\;},
\end{displaymath}
where
\begin{multline*}
\delta{Q_{x}} = \frac{m_{n}}{\hbar}\cdot\left\{
\frac{1}{L_{1}^{2}}\cdot
\left(v_{i}^{2}+v_{f}^{2}\cdot\frac{L_{2}}{L_{3}}\cdot\cos{2}\theta
\right)^{{\!}{\!}2}\cdot\delta{t_{p}^{2}}+
\frac{1}{L_{1}^{2}}\cdot
\left(v_{i}^{2}+v_{f}^{2}\cdot\frac{L_{1}+L_{2}}{L_{3}}\cdot\cos{2}\theta
\right)^{{\!}{\!}2}\cdot\delta{t_{c}^{2}}\right.\\
\left.+\left(\frac{v_{f}^{2}}{L_{3}}
\cdot\cos{2}\theta\right)^{{\!}{\!}2}\cdot\delta{t_{d}^{2}} +
 \left(v_{f}\cdot\sin{2}\theta\right)
^{2}\cdot\left(\delta{2}\theta\right)^{2}
\right\}^{1/2}
\end{multline*}
and
\begin{multline*}
\delta{Q_{y}} = \frac{m_{n}}{\hbar}\cdot\left\{
\left(\frac{v_{f}^{2}L_{2}}{L_{1}L_{3}}
\cdot\sin{2}\theta\right)^{{\!}{\!}2}\cdot\delta{t_{p}^{2}}+
\left(\frac{v_{f}^{2}}{L_{1}}\cdot\frac{L_{1}+L_{2}}{L_{3}}\cdot\sin{2}\theta
\right)^{{\!}{\!}2}\cdot\delta{t_{c}^{2}}\right.\\
\left.+\left(\frac{v_{f}^{2}}{L_{3}}
\cdot\sin{2}\theta\right)^{{\!}{\!}2}\cdot\delta{t_{d}^{2}} +
\left(v_{f}\cdot\cos{2}\theta\right)
^{2}\cdot\left(\delta{2}\theta\right)^{2}
\right\}^{1/2}
\end{multline*}
The expressions for the energy- and $Q$-resolution yield values that are very consistent with the McStas model, see Figs.~\ref{inelreso} and \ref{momreso}.
Thus they may be used in the scientific analysis of measurements of single crystal excitations such as spin waves or phonons.
On a time of flight instrument, such measurements are commonly accomplished by rotating the crystal around the vertical axis in the laboratory coordinate system, thus changing the direction of the incident neutron wave vector $k\nsub{i}$ in the reciprocal lattice of the sample.
This method was first used by Squires and Pynn to measure phonons in a single crystal
of Mg~\cite{tof1966,tof1972}.
The technique is now routine at CNCS, using a cryostat stick that allows sample rotation while the cryostat itself is stationary.
Currently the implementation is still classic, that is, a separate data file is measured for each sample angle.
Since the instrument data acquisition system saves neutrons as events, it is possible, however, to perform the measurements with {\em continuous} sample rotation, and to correlate the scattered neutrons with the sample orientation after the measurement has been completed.
This provides better coverage of reciprocal space, and it is anticipated that during the coming months, single crystal measurements at CNCS will start using continuous sample rotation.

\section{Conclusion}
\label{Conclusion}

Looking back over a little more than two years of user operation, it can be concluded that CNCS offers excellent intensity and resolution for inelastic and quasielastic neutron scattering experiments in the thermal and cold energy ranges.
A wide range of scientific fields has already been covered, most notably in studies of collective excitations in single crystal samples (spin waves~\cite{laco,feng} and phonons~\cite{pbte}), magnetic nanoparticles~\cite{coo}, the phonon density of states in a thermoelectric
material~\cite{ivo}, and diffusive processes in soft matter.
The most important improvements expected over the next few years will be concerning the instrument background and the overall flexibility with which experiments can be accommodated that require non-standard sample environment equipment.

\section{Acknowledgements}
\label{Acknowledgements}

The following people have made significant contributions to the planning, design or construction of CNCS:
M. C. Aronson,
T. L. Arthur,
W. B. Bailey,
A. Barzilov,
D. A. Bunch,
S. M. Chae,
S. H. Chen,
R. W. Connatser,
J. Cook,
R. G. Cooper,
R. K. Crawford,
R. A. Dean,
R. Dimeo,
W. R. Fox,
F. X. Gallmeier,
M. H. Gevers,
G. E. Granroth,
G. C. Greene,
W. M. McHargue,
L. L. Jacobs,
G. H. Jones,
W. S. Keener,
J. Morgan,
D. Narehood,
N. Page,
A. A. Parizzi,
J. Pearce,
D. B. Prieto,
F. S. Proffitt,
R. A. Riedel,
S. M. Rogers,
C. A. Schnell,
J. L. Stockton,
H. Strauss,
H. Taub,
B. M. Thibadeau,
D. M. Williams,
F. R. Williams,
P. A. Wright.
Construction of CNCS was funded by DOE grant DE-FG02-01ER45912.
Research at Oak Ridge National Laboratory's Spallation Neutron Source was sponsored by the Scientific User Facilities Division, Office of Basic Energy Sciences, U. S. Department of Energy.
We thank E. Pomjakushina and K. Conder for the HoBa$_2$Cu$_3$O$_7$ sample preparation.

\pagebreak

\pagebreak

\pagebreak

\pagebreak

\pagebreak

\pagebreak

\pagebreak

\end{document}